\documentclass[12pt]{article}
\usepackage{amssymb}

\def\thefootnote{\fnsymbol{footnote}}

\newcommand {\ee}{\end{equation}}
\newcommand {\bea}{\begin{eqnarray}}
\newcommand {\eea}{\end{eqnarray}}
\newcommand {\nn}{\nonumber \\}

\newcommand {\pl}{\partial}

\newcommand {\Ga}{\Gamma}

\newcommand {\la}{\lambda}

\newcommand {\del}  {\delta}

\newcommand {\half}{ {\frac{1}{2}} }

\newcommand {\Lcal}{{\cal L}}

\newcommand {\Dcal}{{\cal D}}
\newcommand {\Ncal}{{\cal N}}

\def\overleftarrow#1{\vbox{\ialign{##\crcr
 $\leftarrow$\crcr\noalign{\kern-1pt\nointerlineskip}
 $\hfil\displaystyle{#1}\hfil$\crcr}}}

\newcommand {\labar}{{\bar \lambda}}
\newcommand {\psibar}{{\bar \psi}}

\newcommand {\sibar}{{\bar \sigma}}

\newcommand {\q}    {\quad}
\newcommand {\qq}   {\quad\quad}

\usepackage{graphicx}
\usepackage{latexsym}
\begin{document}
\begin{flushright}
June 2006
\end{flushright}

\vspace{0.5cm}

\begin{center}

{\Large\bf 
The $\del(0)$ Singularity in the Warped Mirabelli-Peskin Model}

\vspace{1.5cm}
{\large Shoichi ICHINOSE
         \footnote{
E-mail address:\ ichinose@u-shizuoka-ken.ac.jp
                  }
}\ and\ 
{\large Akihiro MURAYAMA$^\ddag$
         \footnote{
E-mail address:\ edamura@ipc.shizuoka.ac.jp
                  }
}
\vspace{1cm}

{\large 
Laboratory of Physics, 
School of Food and Nutritional Sciences, 
University of Shizuoka, 
Yada 52-1, Shizuoka 422-8526, Japan
 }

$\mbox{}^\ddag${\large
Department of Physics, Faculty of Education, Shizuoka University,
Shizuoka 422-8529, Japan
}
\end{center}

\vspace{2cm}

\begin{center}
{\large Abstract}
\end{center}

The Mirabelli-Peskin model is a 5D super-Yang-Mills theory
compactified on an orbifold $S^1/Z_2$ with the 4D Wess-Zumino model localized on the boundaries (or branes). As the 5D
gauge multiplet couples to 4D chiral multiplets through delta functions, the model contains singular terms proportional to
$\delta(0)$ after integrating out a 5D auxiliary field. This belongs to the same type of singularity as what was first
noticed by Horava in the orbifold compactification of heterotic string theory. Mirabelli-Peskin showed that this
singularity was field-theoretically harmless by demonstrating its neat cancellation by the singularity produced by the
infinite sum of Kaluza-Klein (KK) excitation modes of bulk propagator. In this paper, the similar cancellation is proved to
occur also in a warped version of Mirabelli-Peskin model with the background of $\mbox{AdS}_5$. The bulk propagator
of scalar component of 5D vector supermultiplet in the warped extra dimension is explicitly KK expanded. Then, its second
derivatives by the coordinates of $S^1/Z_2$ generate a term proportional to $\delta(0)$ at the boundaries. The
cancellation is considered to take place perturbatively to all orders of coupling constants as well as to all loops.

\newpage
\renewcommand{\thefootnote}{\arabic{footnote}}
\setcounter{footnote}{0}

\noindent {{\bf 1}\q{\bf Introduction}}\\

In order to establish a new particle theory beyond the standard model, it is important to construct a consistent
higher-dimensional field theory for the bulk-boundary system. Especially, five-dimensional (5D) field theories with the warped
space-time background like the model by Randall-Sundrum \cite{RS} are very interesting from the viewpoint of, e.g.,
hierarchy problem. In the orbifold construction, as the Lagrangian for localized fields on the boundary is embedded into the bulk
in terms of the delta function, there usually appears delta function squared which in turn produces a singularity proportional to
$\delta(0)$. It was first noticed by Horava \cite{Hor} in the orbifold compactification of heterotic string theory. 

The Mirabelli-Peskin (MP) model \cite{Mirab} is a 5D super-Yang-Mills theory compactified on an orbifold
$S^1/Z_2$ with the four dimensional (4D) Wess-Zumino model localized on the boundaries (or branes). As the 5D gauge
multiplet couples to the 4D chiral multiplets through delta functions, the model contains a singular term of four-body
interaction of boundary chiral scalars proportional to $\delta(0)$ after integrating out a 5D auxiliary field $X^3$. MP
showed that this singularity was field-theoretically harmless by explicitly demonstrating its neat cancellation by the
singularity produced by the infinite sum of Kaluza-Klein (KK) excitation modes of bulk propagator for some specific processes.

The mechanism of such cancellation is as follows: The scalar component ${\mit\Phi}$ of 5D vector multiplet interacts
with the boundary scalars through the derivative coupling. Therefore, the second derivative of the propagator in terms of the
coordinates of $S^1/Z_2$ comes into play in computations of amplitudes for the bulk-boundary system. Then, the infinite sum
of KK excitation modes of bulk propagator generates a term proportional to $\delta(0)$ at the boundaries. This singular term
exactly corresponds to the four-body singular interaction of boundary scalars since this interaction is what will remain
if the propagator is removed. Eventually, the oppositeness of the sign of these two singularities completes the
cancellation. 

The above is on the flat space-time background. For the warped background like $\mbox{AdS}_5$, it is not necessarily
self-evident for the cancellation of $\delta(0)$ singularities to be realized or not. In this paper, we examine if a warped
version \cite{Ichi-Mura-PL4} of MP model can be field-theoretically consistent by explicitly evaluating the second derivatives
of KK expanded propagator of ${\mit\Phi}$. \\  

\noindent{{\bf 2}\q {\bf $\del(0)$ in the flat Mirabelli-Peskin model }}\\

In the MP model, the fifth dimension $y$ is compactified on the orbifold $S^1/Z_2$ of size $l=\pi R$. We have
two 3-branes at the orbifold fixed points $y=0, y=l$. The 5D gauge supermultiplet in the bulk consists of a vector field
$A^M\ (M=0,1,2,3,y)$,  a scalar field $\mit\Phi$, a doublet of symplectic Majorana fields $\la^i\ (i=1,2)$ and a triplet of
auxiliary scalar fields $X^a\ (a=1,2,3)$. All bulk fields are of the adjoint representation of the gauge group. In the following,
we choose it to be U(1). The generalization to other gauge group is straightforward. We project out $\Ncal=1$ supersymmetry
(SUSY) multiplets by assigning $Z_2$-parity to all fields such that $P=+1$ for $A_\mu \ (\mu=0,1,2,3), \la^1_L, X^3$ and
$P=-1$ for $A_y, {\mit\Phi}, \la^2_L, X^1, X^2$, in consistent with the 5D SUSY. Then,
$V\equiv(A_\mu,\la^1_L, D) \ (D\equiv X^3-\pl_y{\mit\Phi})$ and
${\mit\Sigma}\equiv({\mit\Phi} + iA_y,-i\sqrt{2}\la^2_R,X^1 + iX^2)$ constitute an $\Ncal =1$ vector supermultiplet
in Wess-Zumino gauge and a chiral scalar supermultiplet, respectively. 

We introduce 4D massless chiral supermultiplets $S \equiv (\phi,\psi,F)$ and $S' \equiv (\phi ',\psi',F')$ on the brane at $y=0$
and $y=l$, respectively, where $\phi,\phi'$ stand for complex scalar fields, $\psi,\psi'$ Weyl spinors and
$F, F'$ auxiliary fields of complex scalar. 

The 5D action is written as
\begin{eqnarray}
S=\int d^4x\int_{-l}^{l} dy\{\Lcal_{bulk}+\del(y)\Lcal_{brane}+\del(y-l){\Lcal'}_{brane} \}, \label{mp1}
\end{eqnarray}
where $\Lcal_{bulk}$ is a 5D bulk Lagrangian;
\begin{eqnarray}
\Lcal_{blk}= - \half \left[\half {F_{MN}}^2 + (\pl_M{\mit\Phi})^2 + i\labar^i\gamma^M \pl_M\lambda^i + (X^a)^2 +
g\labar^i{\mit\Phi}\la^i\right],  \label{mp2}
\end{eqnarray}
and $\Lcal_{brane}(\Lcal'_{brane})$ denotes a 4D boundary
Lagrangian on the brane at $y=0 \ (y=l)$;
\begin{eqnarray}
\Lcal_{brane} &=& -\Dcal_\mu\phi^\dag \Dcal^\mu\phi-i\psi^\dag \sibar^\mu\Dcal_\mu \psi+F^\dag F
+\sqrt{2}ig(\psibar\labar_L\phi-\phi^\dag\la_L\psi) + g\phi^\dag D\phi, \\
\Lcal'_{brane} &=& \Lcal_{brane}[\phi,\psi,F \longrightarrow \phi ',\psi',F'],
\end{eqnarray}
with the covariant derivative $\Dcal_\mu \equiv \pl_\mu + igA_\mu$.

The scalar field $\phi \ (\phi')$ on the brane at $y=0 \ (y=l)$ couples to $X^3$ through the delta
function in (\ref{mp1}). As $X^3$ is a 5D auxiliary field, it is integrated out to leave on the brane a derivative coupling of
$\phi\ (\phi')$ to the bulk scalar field ${\mit\Phi}$ together with a singular four-$\phi \ (\phi')$ interaction term proportional
to $\del(0)$:
\begin{eqnarray}
&&\int d^4x \int_{-l}^{l}dy \left[\del(y)\{-g\phi^\dag(\pl_y{\mit\Phi})\phi-g^2(\phi^\dag\phi)^2\del(0)\}\right.\nn
&&\hspace{3cm}+\left.\del(y-l)\{-g\phi'^\dag(\pl_y{\mit\Phi})\phi' - g^2(\phi'^\dag\phi')^2\del(0)\}\right].
\end{eqnarray}
MP explicitly demonstrated the neat cancellation of this singularity by the singularity produced through the
infinite sum of KK excitation modes of bulk propagator. 

The mechanism of this cancellation is as follows: If Feynman diagrams of any process include the singular four-4D scalar
vertex (Fig.1(a)) as a building block of the diagram on the brane, we necessarily have a corresponding diagram with the
propagator of ${\mit\Phi}$ inserted in the vertex (Fig.1(b)). The contribution of Fig.1(b) to the amplitude contains a term
proportional to $\del(0)$ which cancels the the $\del(0)$ of Fig.1(a).

\begin{figure}
\begin{center}
\includegraphics[width=100.mm,height=60.55mm]{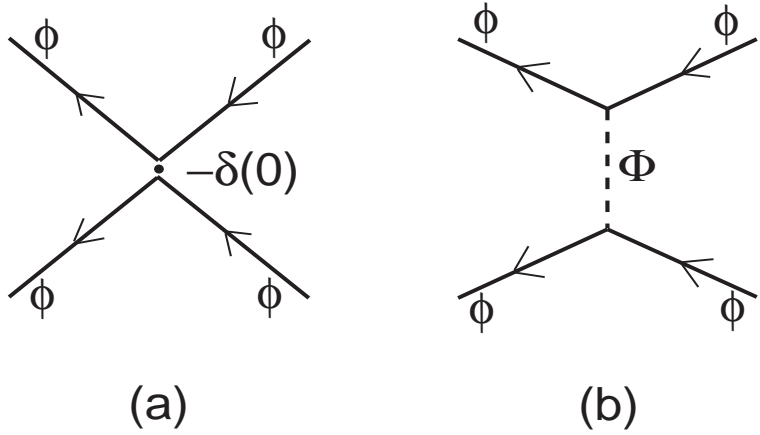}
\end{center}
\caption{Fragments of Feynman diagram. (b) is equivalent to (a) with the vertex replaced by the propagator.}
\label{Fig1}
\end{figure}

We will work in the momentum space for the 4D part, but in the position space for the fifth dimension. We define the
${\mit\Phi}$-propagator $G^{flat}_p(y,y')$ through the 4D Fourier transform of the Green function
\begin{eqnarray}
G^{flat}(x,y; x',y') = \int\frac{d^4 p}{(2\pi)^4}e^{ip(x-x')}G^{flat}_p(y,y'), \label{eq005}
\end{eqnarray}
where $p$ is the four-momentum and $G^{flat}_p(y,y')$ satisfies 
\begin{eqnarray}
[\pl^2_y-p^2]G^{flat}_p(y,y') = \del(y - y'). \label{eq6}
\end{eqnarray}
Suppose the homogeneous solutions are given by 
\begin{eqnarray}
G^{flat}_p(y,y') = A\cosh py + B\sinh py = C\cosh py' + D\sinh py', \label{eq7}
\end{eqnarray} 
for  $0 \leqq y \leqq y' \leqq l$. Since ${\mit\Phi}$ is $Z_2$-odd, (\ref{eq7}) must obey the Dirichlet boundary conditions:
\begin{eqnarray}
G^{flat}_p(0,y') = G^{flat}_p(y, l) = 0. \label{eq8}
\end{eqnarray}
Then, by matching the two solutions over the delta function, we obtain
\begin{eqnarray}
G^{flat}_p(y,y') = \frac{\sinh py \sinh p(y'-l)}{4p\sinh pl}. \label{eq9}
\end{eqnarray}

Making the Fourier expansion of the propagator (\ref{eq9}), we obtain
\begin{eqnarray}
\tilde{G}^{flat}_p(y,y') = \frac{1}{2l}\sum^\infty_{n=1}\frac{f_n(y)f_n(y')}{p^2 + m_n^2}, \label{eq12}
\end{eqnarray}
where 
\begin{eqnarray}
f_n(y) = \sqrt{2}\sin(m_n y), \q (n = 1, 2, \cdots),
\end{eqnarray}
are the KK eigenmodes of ${\mit\Phi}$ for the eigenvalues (KK masses) $m_n = n\pi/l$. $G^{flat}_p(y,y')$ and
$\tilde{G}^{flat}_p(y,y')$ coincide for $0 \leqq y \leqq y' \leqq l$. The propagator (\ref{eq12})  is valid also for negative $y,y'$
and satisfies a generalization of (\ref{eq6}):
\begin{eqnarray}
[\pl^2_y-p^2]\tilde{G}^{flat}_p(y,y') = \half[\del(y - y') - \del(y + y')], \label{eq11}
\end{eqnarray}
which is $Z_2$-odd as well as $y \leftrightarrow y'$-symmetric. We have
\begin{eqnarray}
&&\mbox{\it eigenvalue equation}: \ [\pl^2_y+m_n^2]f_n(y) = 0, \\
&&\mbox{\it boundary condition}: \ f_n(0) = f_n(\pm l) = 0,\\
&&\mbox{\it KK expansion}: \ {\mit\Phi}(x^\mu,y) = \sum_{n=1}^\infty {\mit\Phi}_n(x^\mu)f_n(y), \\
&&\mbox{\it orthonormal condition}: \ \frac{1}{2l}\int_{-l}^l f_n(y)f_m(y) dy = \del_{nm}, \\
&&\mbox{\it completeness}: \ \frac{1}{2l}\sum_{n=1}^\infty f_n(y)f_n(y') = \half[\del(y - y') - \del(y + y')].
\end{eqnarray}
The propagator (\ref{eq005}) with (\ref{eq12}) coincides with the MP's propagator
\begin{eqnarray}
<x,y;x',y'> = \int\frac{d^4 p}{(2\pi)^4}e^{ip(x-x')}\frac{1}{2l}\sum^\infty_{n=-\infty}\frac{e^{i\frac{n\pi}{l}(y-y')} -
e^{i\frac{n\pi}{l}(y+y')}}{p^2 + (\frac{n\pi}{l})^2},
\end{eqnarray}
up to a factor 1/2.

The bulk field ${\mit\Phi}$ interacts with the boundary scalars through the
derivative coupling. Therefore, in computations of amplitudes of the process including the diagram (Fig.1(b)), the
derivative of the propagator such as 
\begin{eqnarray}
\pl^2\tilde{G}^{flat}_p(y, y')/\pl y\pl y' = \frac{1}{l}\sum^\infty_{n=1}\frac{m_n^2\cos(m_n y)\cos(m_n y')}
{p^2 + m_n^2},
\end{eqnarray}
appears, so that
\begin{eqnarray}
\pl^2\tilde{G}^{flat}_p(y, y')/\pl y\pl y'\left|_{y=y'=0}\right. & = & \pl^2\tilde{G}^{flat}_p(y, y')/
\pl y\pl y'\left|_{y=y'=\pm l}\right. \nn
& = & \frac{1}{2l}\sum^\infty_{n=-\infty}\frac{n^2}{n^2 + (\frac{pl}{\pi})^2} = \del(0) - 
\frac{p}{2}\coth pl, \label{eq15}\\
\pl^2\tilde{G}^{flat}_p(y, y')/\pl y\pl y'\left|_{y=0, y'=\pm l}\right. & = & \frac{1}{2l}\sum^\infty_{n=-\infty}
\frac{(-1)^n n^2}{n^2 + (\frac{pl}{\pi})^2} = - \frac{p}{2}\mbox{cosech} pl,
\end{eqnarray}
where we have used the facts
\begin{eqnarray}
\frac{1}{2l}\sum^\infty_{n=-\infty}1 = \del(0), \label{eq16}
\end{eqnarray}
and
\begin{eqnarray}
\sum^\infty_{n=1}(-1)^n = 2l\del(l) = 0.
\end{eqnarray} 
The singularity $\del(0)$ in (\ref{eq15}) is just what cancels that of Fig.1(a).

Notice that the propagator (\ref{eq9}), which is not yet KK expanded, cannot reproduce $\delta(0)$ as it stands. \\

\noindent{{\bf 3}\q {\bf $\del(0)$ in the warped Mirabelli-Peskin model }}\\

We take an anti-deSitter space metric as \cite{RS}
\begin{eqnarray}
ds^2=e^{-2\sigma}\eta_{\mu\nu} dx^\mu dx^\nu + dy^2, \label{AdS01}
\end{eqnarray}
with
\begin{eqnarray}
\sigma = k|y|, \label{eq026}
\end{eqnarray}
where $1/k$ is the AdS$_5$ curvature radius and $\eta_{\mu\nu} = (-1, 1, 1, 1)$ is the 4D metric. We have two 3-branes:
a "Planck brane" at $y=0$ and a "TeV brane" at $y=l$. It is assumed that $k \lesssim M_{Pl}$(the Planck mass) and $kl
\approx 35$ in order that the effective mass scales associated with the "Planck" and "TeV" branes are the Planck scale
and the TeV scale respectively. 

The 5D action is given by (\ref{mp1}) with the following bulk and brane Lagrangians;
\begin{eqnarray}
\Lcal_{blk} &= & - \half \sqrt{-G}\left[\half {F_{MN}}^2 + (\pl_M{\mit\Phi})^2 + i\labar^i\gamma^M \nabla_M\lambda^i +
(X^a)^2 + g\labar^i{\mit\Phi}\la^i \right. \nn
&& \hspace*{6cm} + \left. m_{\mit\Phi}^2{\mit\Phi}^2 + im_\la{\labar}^i(\sigma^3)^{ij}\la^j\right],   \label{eq19}\\
\Lcal_{brane} &=& \sqrt{-G}[-\Dcal_\mu\phi^\dag \Dcal^\mu\phi-i\psi^\dag \sibar^\mu\Dcal_\mu \psi+F^\dag F
 \nn
&& \hspace*{5cm}+\sqrt{2}ig(\psibar\labar_L\phi-\phi^\dag\la_L\psi) + g\phi^\dag D\phi], \\
\Lcal'_{brane} &=& \Lcal_{brane}[\phi,\psi,F \longrightarrow \phi ',\psi',F'],
\end{eqnarray}
where $G=\det (G_{MN}), \nabla_M = \pl_M + {\mit\Ga}_M$ with ${\mit\Ga}_M$ being the spin connection \cite{{Shuster},
{Gher-Pom}}, $m_{\mit\Phi}^2 = -4k^2 + 2\sigma'', m_\la = \half\sigma'$ and the relevant rescaling of the fields have been
done.

The propagator for ${\mit\Phi}$ with the four-momentum $p$ is defined by (\ref{eq005}) with $flat$ replaced by $warped$
and satisfies
\begin{eqnarray}
\left[e^{4\sigma}\pl_y(e^{-4\sigma}\pl_y)+4k^2 - e^{2\sigma}p^2\right]G^{warped}_p(y,y') =
e^{3\sigma}\del(y-y'),\label{eq22}
\end{eqnarray}
for $0 \leqq y \leqq y' \leqq l$. Suppose the homogeneous solutions of (\ref{eq22}) are given by
\begin{eqnarray}
G^{warped}_p(y,y') & = & \frac{e^{2ky}}{k}\left\{AJ_0(ipe^{ky}/k)+BY_0(ipe^{ky}/k)\right\} \nn
& = & \frac{e^{2ky'}}{k}\left\{CJ_0(ipe^{ky'}/k)+DY_0(ipe^{ky'}/k)\right\}. \label{eq23}
\end{eqnarray}
Since ${\mit\Phi}$ is $Z_2$-odd, (\ref{eq23}) must obey the Dirichlet boundary conditions. Then, by matching
the two solutions over the delta function, we obtain
\begin{eqnarray}
G^{warped}_p(y,y') &=& \frac{\pi}{4}\frac{e^{2k(y+y')}/k}{AD-BC}\left\{AJ_0(ipe^{ky}/k)+BY_0(ipe^{ky}/k)\right\}
\nn && \hspace*{3cm}\times\left\{CJ_0(ipe^{ky'}/k)+DY_0(ipe^{ky'}/k)\right\}, \label{eq24}
\end{eqnarray}
where
\begin{eqnarray}
A = Y_0(ip/k), B = - J_0(ip/k), C = Y_0(ipe^{\pi kR}/k), D = J_0(ipe^{\pi kR}/k).
\end{eqnarray}

It is straightforward to expand the propagator (\ref{eq24}) \'a la Fourier expansion in terms of KK eigenmodes:
\begin{eqnarray}
&&\tilde{G}^{warped}_p(y,y') \nn
& = &\sum^\infty_{n=1}\frac{g_n(y)}{\omega'(\lambda)|_{\lambda=m_n^2}}\left[\int_0^{y'}\xi^{-3}
G^{warped}_p(\xi,y')g_n(\xi)d\xi +\int_{y'}^l\xi^{-3}G^{warped}_p(y',\xi)g_n(\xi)d\xi\right]\nn
& = & \frac{1}{2l}\sum^\infty_{n=1}\frac{g_n(y)g_n(y')}{p^2+m_n^2}, \label{eq25}
\end{eqnarray}
where
\begin{eqnarray}
g_n(y) = \frac{e^{2ky}}{N_n}\left[J_0(\frac{m_n}{k}e^{ky}) + b_0(t)|_{t=m_n}Y_0(\frac{m_n}{k}e^{ky})\right], 
\end{eqnarray}
with
\begin{eqnarray}
b_0(t) &\equiv & -J_0(t/k)/Y_0(t/k), \\
b_0(t)|_{t=m_n} & = & b_0(t)|_{t=m_ne^{kl}}, \label{eq36}\\
N_n^2 & = & \frac{1}{l}\int^l_0 dy e^{2ky}\left[J_0(\frac{m_n}{k}e^{ky}) +
b_0(t)|_{t=m_n}Y_0(\frac{m_n}{k}e^{ky})\right]^2,
\end{eqnarray}
are the KK eigenmodes of ${\mit\Phi}$ for the eigenvalues (KK masses) $m_n$ and
\begin{eqnarray}
\omega(\lambda) \equiv \frac{2k^4}{\pi N_n^2}\left\{b_0(t)|_{t=\sqrt{\lambda}e^{kl}}-b_0(t)|_{t=\sqrt{\lambda}}\right\}.
\end{eqnarray}
$G^{warped}_p(y,y')$ and $\tilde{G}^{warped}_p(y,y')$ coincide for $0 \leqq y \leqq y' \leqq l$.
Namely, we have
\begin{eqnarray}
&&\mbox{\it eigenvalue equation}: \ \left\{e^{4ky}\pl_y(e^{-4ky}\pl_y)+4k^2+e^{2ky}m_n^2\right\}g_n(y)=0,
\label{eq38}\\
&&\mbox{\it boundary condition}: \ \ g_n(0) = g_n(l) = 0, \label{eq39}\\
&&\mbox{\it KK expansion}: \ {\mit\Phi}(x^\mu,y) = \sum_{n=1}^\infty {\mit\Phi}_n(x^\mu)g_n(y), \label{eq40}\\
&&\mbox{\it orthonormal condition}: \ \frac{1}{l}\int_{0}^l e^{-2ky} g_n(y)g_m(y) dy = \del_{nm}, \label{eq41}\\
&&\mbox{\it completeness}: \ \frac{1}{2l}\sum_{n=1}^\infty g_n(y)g_n(y') = e^{2ky}\del(y - y'). \label{eq42}
\end{eqnarray}
We extend the domain of the fifth coordinate to negative values so that the propagator is odd under $Z_2$ by modifying
$g_n(y)$ such that \cite{Gher-Pom}
\begin{eqnarray}
g_n(y) = \epsilon(y)\frac{e^{2\sigma}}{N_n}\left[J_0(\frac{m_n}{k}e^{\sigma}) +
b_0(t)|_{t=m_n}Y_0(\frac{m_n}{k}e^{\sigma})\right], \label{eq43}
\end{eqnarray}
where $\epsilon(y) = \sigma'/k$ is a sign function. The modified eigenfunction (\ref{eq43}) is confirmed to satisfy
(\ref{eq38}) $\sim$ (\ref{eq42}) with $ky$ replaced by $\sigma$ (\ref{eq026}).  Then, making use of the fact
\begin{eqnarray}
\epsilon(y)\epsilon(y')=\half\{\epsilon(y)\epsilon(y')-\epsilon(y)\epsilon(-y')\},
\end{eqnarray}  
the right-hand side of (\ref{eq42}) is naturally extended so as to include a term $\propto \del(y+y')$:
\begin{eqnarray}
\frac{1}{2l}\sum_{n=1}^\infty g_n(y)g_n(y') = \frac{e^{2\sigma}}{2}\{\del(y - y')-\del(y+y')\},
\end{eqnarray}
which gives
\begin{eqnarray}
&&\left[e^{4\sigma}\pl_y(e^{-4\sigma}\pl_y)+4k^2 - e^{2\sigma}p^2\right]\tilde{G}^{warped}_p(y,y') \nn
&& \hspace*{6cm} = \frac{e^{3\sigma}}{2}\left\{\del(y-y')-\del(y+y')\right\}, \label{eq047}
\end{eqnarray}
so that the $Z_2$-parity is properly taken into account.

Fig.1(b) corresponds to the second derivative of the propagator (\ref{eq25}) in terms of $y$ and $y'$:
\begin{eqnarray}
\pl^2\tilde{G}^{warped}_p(y, y')/\pl y\pl y' = \frac{1}{2l}\sum^\infty_{n=1}\frac{g'_n(y)g'_n(y')}{p^2+m_n^2}. \label{eq47}
\end{eqnarray} 
We first estimate the quantity $g'_n(y)g'_n(y')$ on the branes as follows:
\begin{eqnarray}
&& g'_n(y)g'_n(y')|_{y=y'=0} \ = \frac{2klm_n^2}{Y_0^2(m_n/k)/Y_0^2(m_n e^{kl}/k)-1}, \label{eq48}\\
&& g'_n(y)g'_n(y')|_{y=y'=l} =  \frac{2klm_n^2}{1-Y_0^2(m_n e^{kl}/k)/Y_0^2(m_n/k)},\label{eq49}\\
&& g'_n(y)g'_n(y')|_{y=0, \ y'=l} \nn
&& \hspace*{1.5cm} = \frac{2klm_n^2}{Y_0(m_n/k)/Y_0(m_n e^{kl}/k)-Y_0(m_n e^{kl}/k)/Y_0(m_n/k)}e^{3kl}.
\label{eq50}
\end{eqnarray}

The KK masses are given by solutions of the condition (\ref{eq36}). We assume $kl \gg 1$ and solve it for three different
ranges of $m_n$. Namely, we obtain
\begin{eqnarray}
\mbox{i)}&& m_n \ \approx n\pi ke^{-kl}, \q\mbox{for}\q m_n \gg k,\\
\mbox{ii)}&& m_n \ \approx \left(n+\frac{1}{4}\right)\pi ke^{-kl}, \q\mbox{for}\q m_n \approx k,\\ 
\mbox{iii)}&& m_n \ \approx \left(n-\frac{1}{4}\right)\pi ke^{-kl}, \q\mbox{for}\q m_n \ll k,
\end{eqnarray}
which give an approximation of the ratio $Y_0(m_n/k)/Y_0(m_n e^{kl}/k)$ as follows;
\begin{eqnarray}
&& Y_0(m_n/k)/Y_0(m_n e^{kl}/k) \approx (-1)^n e^{kl/2}, \q\mbox{for}\q m_n \gtrsim k,\\
&& Y_0(m_n e^{kl}/k)/Y_0(m_n/k) \approx 0, \q\mbox{for}\q m_n \ll k.
\end{eqnarray}
Then, (\ref{eq48})$\sim$(\ref{eq50}) can be estimated in a good approximation as shown in Table I.
\begin{table}
\caption{$g'_n(y)g'_n(y')$ for various limits.}
\label{tab1}
\begin{center}
\begin{tabular}{|l|c|c|}
\hline
 & $m_n \gtrsim k$ & $m_n \ll k$ \\
\hline\hline
$g'_n(y)g'_n(y')|_{y=y'=0}$              & $\approx 2kle^{-kl}m_n^2$ & $\approx 0$\\
\hline
$g'_n(y)g'_n(y')|_{y=y'=l}$ & $\approx 2klm_n^2$ & $\approx 2klm_n^2$ \\
\hline
$g'_n(y)g'_n(y')|_{y=0, \ y'=l}$ & $\approx 2kl(-1)^ne^{-5kl/2}m_n^2$ & $\approx 0$ \\ 
\hline 
\end{tabular}
\end{center}
\end{table}

On the "Planck" brane, i.e., $y=y'=0$, high momentum ($p\gtrsim k$) processes will be dominant and contributions of $m_n$
much lower than $k$ ($m_n \ll k$) to the summation  in (\ref{eq47}) are negligible. Therefore, we obtain
\begin{eqnarray}
\pl^2\tilde{G}^{warped}_p(y, y')/\pl y\pl y'|_{y=y'=0} &\approx & ke^{-kl}\sum^\infty_{n=1}1 - \half p\coth(pe^{kl}/k). \label{eq56}
\end{eqnarray}
By making use of such a representation of $\displaystyle{\sum^\infty_{n=1}1}$ as
\begin{eqnarray}
\sum^\infty_{n=1}1 = \left(e^{akl}+\half\right)\del(0)/k + 2e^{akl}, \label{eq59}
\end{eqnarray}
with $a=1$, we find
\begin{eqnarray}
\pl^2\tilde{G}^{warped}_p(y, y')/\pl y\pl y'|_{y=y'=0} \approx \del(0) + 2k - \half p\coth(pe^{kl}/k). \label{eq60}
\end{eqnarray}
It is the singular term $\propto \del(0)$ in (\ref{eq60}) that cancels $\del(0)$ of Fig.1(a). 

On the "TeV" brane, i.e., $y=y'=l$, Fig.1(b) represents
\begin{eqnarray}
\pl^2\tilde{G}^{warped}_p(y, y')/\pl y\pl y'|_{y=y'=l} \approx k\sum^\infty_{n=1}1 - \half pe^{kl}\coth(pe^{kl}/k), \label{eq57}
\end{eqnarray}
multiplied by the factor of metric at the vertices. In order for the singularity in it to cancel the contribution of Fig.1(a), we
use (\ref{eq59}) with $a=4$ and obtain
\begin{eqnarray}
\pl^2\tilde{G}^{warped}_p(y, y')/\pl y\pl y'|_{y=y'=l} &\approx & - e^{4kl}[\del(0) + 2k] - \half pe^{kl}\coth(pe^{kl}/k).
\end{eqnarray}

Notice that the derivative of Fourier expansion of a linear combination of sign function and sawtooth-wave function
\cite{Ichi-Mura-PL3} such as
\begin{eqnarray}
F(x,k,a) \equiv \frac{1}{\pi k}\sum^\infty_{n=1}\left[(4e^{akl}+1)\frac{\sin\{(2n-1)\pi kx\}}{2n-1}- e^{akl}\frac{\sin(4n\pi
kx)}{n}\right],
\label{eq61}
\end{eqnarray}
gives (\ref{eq59}) in the limit $x \rightarrow 0$. Indeed, 
\begin{eqnarray}
\frac{\pl F(x,k,a)}{\pl x} = \sum^\infty_{n=1}\left[(4e^{akl}+1)\cos\{(2n-1)\pi kx\} - 4e^{akl}\cos(4n\pi kx)\right].
\label{eq71}
\end{eqnarray}
If we take the limit $x \rightarrow 0$ before the summation in the right-hand side of (\ref{eq71}), we have
\begin{eqnarray}
\left. \frac{\pl F(x,k,a)}{\pl x}\right |_{x=0} =  \sum^\infty_{n=1} 1. \label{eq64}
\end{eqnarray}
If we carry out the summation first, then we obtain
\begin{eqnarray}
\left. \frac{\pl F(x,k,a)}{\pl x}\right |_{x=0} = \left(e^{akl}+\half\right)\del(0)/k + 2e^{akl}. \label{eq650}
\end{eqnarray}
(\ref{eq64}) and (\ref{eq650}) reproduce (\ref{eq59}). 

If ${\mit\Phi}$ propagates from the "Planck" brane to the "TeV" brane, i.e., $y=0,\ y'=l$, no singularity appears. Namely, we
have
\begin{eqnarray}
&& \pl^2\tilde{G}^{warped}_p(y, y')/\pl y\pl y'|_{y=0, \ y'=l} \nn
&& \hspace{2cm}\approx ke^{-kl/2}\sum^\infty_{n=1}(-1)^n - \half pe^{3kl/2}\mbox{cosech}(pe^{kl}/k), \nn
&& \hspace{2cm}\approx - \half pe^{3kl/2}\mbox{cosech}(pe^{kl}/k). \label{eq58}
\end{eqnarray} 

It should be remarked that the propagator (\ref{eq24}), which is not yet KK expanded, cannot reproduce $\delta(0)$ as it
stands.
\\

\noindent{{\bf 4}\q {\bf The flat limit of the warped Mirabelli-Peskin model }}\\

Expansions such as
\begin{eqnarray}
J_0(\frac{m_n}{k}e^{k|y|}) &=& J_0(\frac{m_n}{k}) - m_nJ_1(\frac{m_n}{k})|y| + \half
m_n^2\left\{-J_0(\frac{m_n}{k})+\frac{k}{m_n}J_1(\frac{m_n}{k})\right\}|y|^2 \nn
&+& \frac{1}{6}m_n^3\left[\frac{k}{m_n}J_0(\frac{m_n}{k})+\left\{1-2(\frac{k}{m_n})^2\right\}J_1
(\frac{m_n}{k})\right]|y|^3 + \cdots,\nonumber\\  
&\stackrel{k\rightarrow 0}{\longrightarrow}& J_0(\frac{m_n}{k})\cos (m_n |y|) -
J_1(\frac{m_n}{k})\sin(m_n|y|),\label{eq65} 
\end{eqnarray}
\begin{eqnarray}
Y_0(\frac{m_n}{k}e^{k|y|}) = (J\rightarrow Y) \qq \mbox{in (\ref{eq65})},
\end{eqnarray}
give
\begin{eqnarray}
&&Y_0(\frac{m_n}{k})J_0(\frac{m_n}{k}e^{k|y|}) - J_0(\frac{m_n}{k})Y_0(\frac{m_n}{k}e^{k|y|}) \nn
& = & \left\{J_0(\frac{m_n}{k})Y_1(\frac{m_n}{k}) -
J_1(\frac{m_n}{k})Y_0(\frac{m_n}{k})\right\}\left\{m_n|y|-\frac{1}{3!}(m_n|y|)^3+\cdots\right\} \nn
&&  + \ O((k/m_n)^2) \nn
&\stackrel{k\rightarrow 0}{\longrightarrow}& \frac{2k}{\pi m_n}\sin(m_n|y|), \label{eq67}
\end{eqnarray}
and
\begin{eqnarray}
N_n^2 &\stackrel{k\rightarrow 0}{\longrightarrow}& \ \frac{1}{\pi R}\int_0^{\pi R}dye^{2ky}\left[J_0(\frac{m_n}{k})\cos (m_n y)
- J_1(\frac{m_n}{k})\sin(m_n y)\right. \nn
&&-\left.\frac{J_0(\frac{m_n}{k})}{Y_0(\frac{m_n}{k})}\left\{Y_0(\frac{m_n}{k})\cos (m_n y) - Y_1(\frac{m_n}{k})\sin(m_n
y)\right\}\right]^2 \nn
&=&\frac{1}{\pi R}\int_0^{\pi R}dye^{2ky}\left[J_1(\frac{m_n}{k}) - \frac{J_0(\frac{m_n}{k})}{Y_0(\frac{m_n}{k})}
Y_1(\frac{m_n}{k})\right]^2\sin^2(m_ny) \nn
&=&\frac{1}{\pi RY_0^2(\frac{m_n}{k})}\int_0^{\pi R}dye^{2ky}\frac{4k^2}{\pi^2m_n^2}\sin^2(m_ny) \nn
&\approx& \frac{2k^2}{\pi^2m_n^2Y_0^2(\frac{m_n}{k})}, \label{eq68}
\end{eqnarray} 
so that
\begin{eqnarray}
g_n(y) \stackrel{k\rightarrow 0}{\longrightarrow} f_n(y) = \sqrt{2}\sin (m_ny).
\end{eqnarray}

For the sake of (\ref{eq67}) and (\ref{eq68}), the condition (\ref{eq36}) is reduced to
\begin{eqnarray}
2m_n\sin (m_nl) = 0,
\end{eqnarray}
for $k\rightarrow 0$. Therefore, we have
\begin{eqnarray}
m_n \stackrel{k\rightarrow 0}{\longrightarrow} \pi n/l, \qq n = 1, 2, \cdots.
\end{eqnarray}

Finally, if we take the limit $k \rightarrow 0$ before the summation in the right-hand side of (\ref{eq71}), we have
\begin{eqnarray}
\frac{\pl F(x,k,a)}{\pl x} \stackrel{k\rightarrow 0}{\longrightarrow} \sum^\infty_{n=1}1 = \half\sum^\infty_{n=-\infty}1.
\label{eq74}
\end{eqnarray}
If we carry out the summation first, then we obtain
\begin{eqnarray}
\frac{\pl F(x,k,a)}{\pl x} = \left(e^{akl}+\half\right)\del(kx) + 2e^{akl}.
\end{eqnarray} 
Let us introduce dimension-less quantities $\hat{x} \equiv x/l$ and $\hat{k} \equiv lk$. Then, we have
\begin{eqnarray}
\del(kx) = \del\left(\frac{\hat{k}}{l}\hat{x}l\right) = \frac{l}{\hat{x}}\del(l\hat{k}),
\end{eqnarray}
and
\begin{eqnarray}
\left. \frac{\pl F(x,k,a)}{\pl x}\right |_{\hat{x}=3/2, \hat{k}=0} = l\del(0) + 2. \label{eq77}
\end{eqnarray}
Thus, we recover (\ref{eq16}). \\

\noindent{{\bf 5}\q {\bf Concluding remarks}}\\

We have explicitly shown that the cancellation of $\del(0)$ singularities in the MP model is realized not only for
the flat space-time background but also for the warped background of $\mbox{AdS}_5$. Such a singularity originates in
the delta functions which connect the 4D fields with the 5D fields in the orbifold picture. In the MP model, the
four-4D scalar vertex embodies the $\del(0)$ singularity which emerges in the Lagrangian after eliminating the 5D auxiliary
field $X^3$. There necessarily corresponds the diagram with the propagator of 5D scalar ${\mit\Phi}$ inserted in the vertex
point. The contribution of this diagram to the amplitude for any process is proportional to the second derivative of the
propagator along the extra dimension. Then, the infinite sum of KK modes of ${\mit\Phi}$ gives rise to $\del(0)$ on the brane. 

The essentials of above mechanism of cancellation is simple and in common for both backgrounds; flat and warped. Therefore,
we have advanced the expatiation correspondingly and shown explicitly the flat limit of KK eigenmodes, KK masses and the
representation of infinite sum of unity in the warped MP model. It deserves great attention that the propagators (\ref{eq9}) and
(\ref{eq24}), which are not yet KK expanded and represented in the momentum space for the 4D part but in the position space
for the extra dimension, cannot reproduce $\delta(0)$ with respect to Fig.1(b). We should use the KK expanded propagators
(\ref{eq12}) and (\ref{eq25}) in computing Feynman diagrams containing Fig.1(b). It should be remarked, too, that there
appears $\delta(y+y')$ in addition to $\delta(y-y')$ in the right-hand side of equations (\ref{eq11}) and (\ref{eq047}) which
(\ref{eq12}) and (\ref{eq25}) satisfy, respectively.

In the warped MP model, the KK mass is given by the solution of an equation involving the
Bessel functions and cannot be written analytically. Therefore, the extraction of $\del(0)$ from the derivative of the
${\mit\Phi}$-propagator is only approximately done. Now that we have established the cancellation mechanism, we should
rather give priority to the exact cancellation of singularities and develop approximations around it.

Since the lines of 4D scalars in Fig.1 are irrelevant to whether they are external or internal and there is a perfect one-to-one
correspondence between (a) and (b), the cancellation is considered to take place perturbatively to all orders of coupling
constants as well as to all loops for any process.


\begin{thebibliography}{99}
\bibitem{RS} L.Randall and R.Sundrum, Phys.Rev.Letters {\bf 83} (1999) 3370.
\bibitem{Hor} P.Horava, Phys.Rev. {\bf D54} (1996) 7561 [arXiv:hep-th/9608019]
\bibitem{Mirab} E.A.Mirabelli and M.E.Peskin, \newblock Phys.Rev. {\bf D58} (1998) 065002.
\bibitem{Ichi-Mura-PL4} S.Ichinose and A.Murayama, Phys.Letters {\bf 625B} (2005) 106 [arXiv:hep-th/0409193].
\bibitem{Shuster} E.Shuster, Nucl.Phys. {\bf B554} (1999) 198 [arXiv:hep-th/9902129].
\bibitem{Gher-Pom} T.Gherghetta and A.Pomarol, Nucl.Phys. {\bf B586} (2000)141 [arXiv:hep-ph/0003129].
\bibitem{Ichi-Mura-PL3} S.Ichinose and A.Murayama, Phys.Letters {\bf 596B} (2004) 123 [arXiv:hep-th/0405065].
\end{thebibliography}
\end{document}